\documentclass[aps,pre,nofootinbib,twocolumn,superscriptaddress,
  amsmath,amssymb,floatfix,showpacs]{revtex4} 

\usepackage[latin1]{inputenc}
\usepackage[english]{babel}
\usepackage{amsmath}
\usepackage{amssymb}
\usepackage{bm}

\usepackage{makeidx}
\usepackage{pstricks,pst-grad}

\usepackage{graphicx}

\newcommand{\ep}{\epsilon}

\newcommand{\be}{\begin{equation}}
\newcommand{\bea}{\begin{eqnarray}}

\newcommand{\ee}{\end{equation}}
\newcommand{\eea}{\end{eqnarray}}
\newcommand{\belign}{\begin{align}}
\newcommand{\elign}{\end{align}}

\newcommand{\si}{\sigma_{\alpha}}
\newcommand{\la}{\lambda}

\newcommand{\Nm}{\mathcal{N}}

\title{Dynamical Transition in the Directed and REM-like Trap Model}
\begin{document}

\title{Slow relaxation, dynamic transitions and extreme value
  statistics in disordered systems} 

\author{K. van Duijvendijk}

\affiliation{Laboratoire Mati\`ere et Syst\`emes Complexes (CNRS UMR
  7057), Universit\'e de Paris VII, 10 rue Alice Domon et L\'eonie
  Duquet, 75205 Paris Cedex 13, France}

\author{G. Schehr}

\affiliation{Laboratoire de Physique Th\'eorique (CNRS UMR 8627),
 Universit\'e 
 de Paris-Sud, 91405 Orsay Cedex, France}

\author{F. van Wijland $^1$}

\begin{abstract}
We show that the dynamics of simple disordered models, like the
directed Trap
Model and the Random Energy Model, takes place at a coexistence
  point between active and inactive dynamical phases.  
We relate the presence of a dynamic phase transition in these models 
to the extreme value statistics of the associated random energy landscape.
\end{abstract}


\maketitle

\section{Introduction}

The dynamics of glassy systems is by definition out-of-equilibrium
 over experimental time-scales. Glassiness  
 manifests itself through a great variety of dynamical 
 features such as aging, non-exponential relaxation of
 correlation functions or super-Arrhenius slowing down of the
 dynamics \cite{cugl}. However the definition of glassiness remains an open
 problem because, in many situations, no static parameters were found to
 indicate whether
 or not a system is in a glassy state.  There is indeed a general
 agreement about the fact that the glassiness of a system does not necessarily 
 arise from an underlying static transition \cite{krauth}. 

To shed light on this question several dynamical approaches have been
developed \cite{cugl}. Here we explore the suggestion made in \cite{garr} that
glassiness arises in a system when a coexistence between active and inactive
regions of space-time takes place. The idea is that these {\it dynamical
  heterogeneities} are a defining feature of glassy systems. A method
to inquire into the space-time character of a system is the 
application of the thermodynamic formalism of histories developed by Ruelle
and coworkers \cite{ruelle}. While the equilibrium statistical formalism
studies the fluctuations in the configuration space of the system, Ruelle's
formalism focuses on the time realizations the
system can follow in configuration space.
A central parameter in this formalism is the activity $K(t)$ of a
history, which is the number of changes of configurations between
the initial time, set to zero, and time $t$. This parameter is a physical,
time-extensive observable.

For systems without quenched disorder, it was suggested that when for
large times a system is found to have two well 
separated sets of histories -- one where the activity is extensive in the
system size, and the other where the activity is subextensive -- there will be
a coexistence of active and inactive regions of space-time in the system,
separated by sharp interfaces. Thus the system will exhibit slow dynamics.
This was indeed successfully shown in the case of Kinetically Constrained
Models of glasses \cite{lettre} and for the non-equilibrium steady state of
some Markov processes \cite{fred}.

Here we investigate the space-time properties  of a family of disordered
systems. Besides the fact that these exhibit experimental features akin to
those of structural glasses, there is some intrinsic interest to focus on the
effect of quenched disorder. As will become clearer in the next sections, the
central part of our study bears on the statistics of the activity $K(t)$,
which is defined for each realization of disorder. From the probability distribution of $K$ one can
build a mathematical object --its large deviation function-- which in many
respects can be argued to play the role of a dynamical free energy. As is known
from the statics of disordered systems, the presence of quenched disorder
induces additional difficulties when it comes to averaging over the
disorder degrees of freedom. We 
have had to deal with similar ones for the dynamical free energy we are after
in this work.

In this paper we focus on the directed Trap Model and the Random
Energy Model. Although they 
are simpler to study they reproduce experimental features of more
complex disordered and glassy systems like the 
super-activated behavior of the viscosity in glasses \cite{visco}, aging
\cite{trap,revue} or non-trivial violations of the fluctuation-dissipation
theorem~\cite{ritort}. In both cases, we show analytically that the
aforementioned large deviation function displays a discontinuity between the
sets of active  
and inactive histories : this suggests that also in these disordered
systems there is a signature of the glassy behavior using the
thermodynamic formalism of 
histories. Furthermore we study the influence of the
distribution of the disorder on the presence of a dynamic transition
and discuss its occurrence in  
connection \cite{extreme,extr} with the distributions of extrema of
disordered energy landscapes.

\section{Directed Trap Model}

A trap model is defined by $N$ independent traps labeled by an integer $i$,
each 
trap being characterized by an energy $E_i$. 
We consider a continuous time Markov dynamics among traps: the
dynamical evolution is specified by the probability $P_i(t)$ that the
system 
stays in trap $i$ at time $t$, and by the transition rates $W_{ij}$ for jumping
from trap $j$ to trap $i$. In the {\it directed trap model} 
the transition rates take the form 
\be\label{transitionrates}
W_{ij}=\delta_{i,j+1}\frac{1}{B_j} \qquad \text{where} \quad B_j=e^{-\beta
  E_j}\,,  
\ee 
$i=1,\dots ,N$. Thus the system evolves through the following Master Equation:
\be\label{mastertrap}
\frac{dP_i(t)}{dt}=-\frac{P_i(t)}{B_i}+\frac{P_{i-1}(t)}{B_{i-1}}\;.
\ee
The trapping times $B_j$ are random variables
distributed according to a L\'evy distribution
\be\label{def_levy}
p(B_j)=\theta(B_j-1) \mu B_j^{-1-\mu} \qquad, \quad \mu\in(0,1)\,,
\ee
where $\theta(x)$ is the Heaviside step function. In the following we will
use the notation $\overline{{\cal O}(B_j)} = \int p(B_j) {{\cal
    O}(B_j)} dB_j$ to  
denote an average over $B_j$. In particular, the distribution in
Eq. (\ref{def_levy}) implies that the mean trapping time is infinite, which
causes anomalous diffusion.

\subsection{Large Deviation Function}

Ruelle's thermodynamic formalism enables to investigate the
dynamics of a system by studying the histories the system can follow in the
configuration space. In the directed Trap Model configurations are
represented by traps so that a 
history in the configuration space between time $0$ and time $t$ is specified 
by a sequence of traps visited during this time and by the time intervals
elapsed between each jump from one trap to the next. To perform a statistics
over histories one has to classify them according to a 
time-extensive (and history-dependent) parameter. A suitable one is the {\it
  activity} $K(t)$, defined as the number of
configurations, here the number of traps, visited between time zero and
time $t$. In the directed trap model, if the dynamics starts at time
$t=0$ in the trap $i=0$, $K(t)$ is simply the trap
in which the system stands at time $t$. 

We now consider the probability $P_i(K,t)$ to be in trap $i$ at time $t$ at
fixed activity, and define its Laplace transform:
\be\label{def_ptilde}
\tilde{P}_i(s,t)=\sum_{K=0}^\infty e^{-sK}P_i(K,t)\,.
\ee
It can be shown that $\tilde{P}_i(s,t)$ obeys an evolution
equation (similar to but different from a Master Equation) of the form
$\partial_t \tilde{P}_i(s,t)=\hat{W}^K \tilde{P}_i(s,t)$ where the
elements of the evolution operator take the form:
\be\label{def_what}
\hat{W}^K_{ij}=\delta_{j,i-1}\frac{e^{-s}}{B_{i-1}}-\delta_{j,i}\frac{1}{B_{i}}\,. 
\ee

From $\tilde{P}_i(s,t)$ in Eq. (\ref{def_ptilde}), 
one defines the cumulant generating function of the activity $K$:

\be\label{kmoy}
Z_K(s,t)=\sum_{i}\tilde{P}_i(s,t)= \langle e^{-s{ K}}\rangle \;,
\ee
where $\langle \dots \rangle$ stands for an average over all possible
histories the system can follow between time $0$ and time $t$.
At large time $t$ one expects 
\be\label{def_psi}
Z_K(s,t)\sim \,e^{\,t \psi_K(s)} \;,
\ee
where $\psi_K(s)$ is the largest
eigenvalue of the linear operator $\hat{W}_K$ in
Eq. (\ref{def_what}) and is thus a large deviation function. At large
time $t$, the derivatives of $\psi_K(s)$ will give the cumulants of
the activity $K(t)$. Note that 
$\psi_K(s)$ depends a priori on the realization of the random variables
$B_i$ (\ref{def_levy}) and will thus be itself a random variable. 

The parameter $s$ allows to probe the different histories the system can
follow: positive $s$ will correspond to the inactive phase of the dynamics,
{\it i.e.} $K$ smaller than its average, while negative $s$ correspond to the
active phase of the dynamics, {\it i.e.} $K$ larger than its average. A
discontinuity in the derivatives of the large deviation function will 
correspond to a dynamic transition between two different phases (the active
and the inactive one), reflecting the sharp 
interfaces between dynamical heterogeneities.

\subsection{Dynamic Transition}

To compute the large deviation function $\psi_K(s)$ we have to find the
largest eigenvalue of $\hat{W}^K$ defined in Eq. (\ref{def_what}). The
$N$ eigenvalues $\lambda_j(s)$, $1 \leq j \leq N$ are solutions
of the equation 
\be\label{eighen}
e^{-s}\frac{ \tilde{P}_{i-1}(s,t)}{B_{i-1}}-\frac{
  \tilde{P}_i(s,t)}{B_{i}}=\la_j(s)  \tilde{P}_i(s,t)\;,
\ee
and $\psi_K(s) = {\rm max}_{1 \leq j \leq N} \;\{ \lambda_j(s)\}$.
The computation of the characteristic polynomial is straightforward
and leads to the eigenvalue equation 
%
\be\label{eig}
\frac{1}{N}\sum_{i=1}^N \ln(1+\la_j(s) B_i)=-s\,.
\ee
Obviously, one has $\psi_K(s=0)=0$. Let us first consider the case
$s<0$ where there is {\it only one} $\lambda_j > 0$ solution
of Eq. (\ref{eig}) which thus coincides with $\psi_K(s)$. In that
case, {\it i.e.} in the active phase, it is natural to assume that
$\psi_K(s)$ is 
self-averaging so that, in the 
limit $N \to \infty$, one expects
$\psi_K(s)~\simeq~\overline{\psi_K(s)}$. Using the law of large numbers to
treat the sum 
in Eq. (\ref{eig}), one obtains in the $N \to \infty$ limit
\bea\label{exactmean_trap}
\mu \int_1^\infty dB B^{-1-\mu} \ln(1+\overline{\psi_{K}(s)} B)=-s \;,
\eea 
which uniquely determines $\overline{\psi_{K}(s)}$. To describe the
  fluctuations of $\psi_{K}(s)$ around its mean for finite $N \gg 1$
  one writes $\psi_K(s)-\overline{\psi_K(s)} = \chi_K(s) N^{-1/2}
  +\mathcal{O}({N}^{-1})$  
where $\chi_K(s)$ is a zero-mean random variable. Using the Central
Limit Theorem applied to the sum in Eq. (\ref{eig}), one obtains that
$\chi_K(s)$ is a Gaussian variable such that the distribution of
$\psi_K(s)$ is given for $N \gg 1$ by
\be\label{dis_lambda}
p(\psi_K(s))=\frac{\sqrt{N}}{\si\sqrt{
    2\pi}}\exp{\left[-{\frac{N(\psi_K(s)-\overline{\psi_K(s)})^2}{2\si^2}}\right] } 
\ee
where $\si^2$ is given by 
\bea\label{var}
\si^2&=&\left(\overline{\frac{B}{1+ \overline{\psi_K(s)}
      B}}\right)^{-2}\\
&\,&\times[\overline{\ln^2(1+ \overline{\psi_K(s)}
    B)}-(\overline{\ln(1+\overline{\psi_K(s)}   B)})^2]\;,\nonumber 
\eea
where $\overline{\psi_K(s)}$ is given in Eq. (\ref{exactmean_trap}). 

Let us now consider $s > 0$ and label the $B_i$'s such that
$0<B_1<\dots<B_N$. In that case, one obtains the bounds
\be\label{bounds_trap}
-\frac{1}{B_N} < \psi_K (s) < 0 \;.
\ee
From the distribution of the variables $B_i$'s in Eq.~(\ref{def_levy})
one obtains that the one of the largest one $B_N$ is given by:
\be\label{max_dis}
p(B_N)=N \mu B_N^{-1-\mu}(1-B_N^{-\mu})^{N-1} \;,
\ee
from which one gets for $N\gg 1$:
\be
\overline{ -\frac{1}{B_N}} \simeq -
N^{-\frac{1}{\mu}}\Gamma(1+\frac{1}{\mu})e^{-\frac{1}{\mu}+1} \;.
\ee 
Thus, from the bounds in Eq. (\ref{bounds_trap}) we find that
$\lim_{N\rightarrow\infty}\overline{ \psi_K(s) }=0$ for $s > 0$. On
the other hand, for $s < 0$, one has from Eq. (\ref{exactmean_trap})
$\lim_{N\rightarrow\infty} \overline{\psi_K(s)}~>~0$ and in that limit 
$\overline{\psi_K(s)} \propto (-s)^{1/\mu}$ for small $s$. Therefore,
  the first derivative $\overline{\psi'_K(s)}$ is continuous but
  higher order derivatives $\overline{\psi^{(p)}_K(s)}$ with $p \geq
  1/\mu$ will display a discontinuity in $s=0$, indicating a dynamic phase
  transition (of order higher than one).  

\begin{figure}[h]
\begin{centering}
\includegraphics[width=7.7cm]{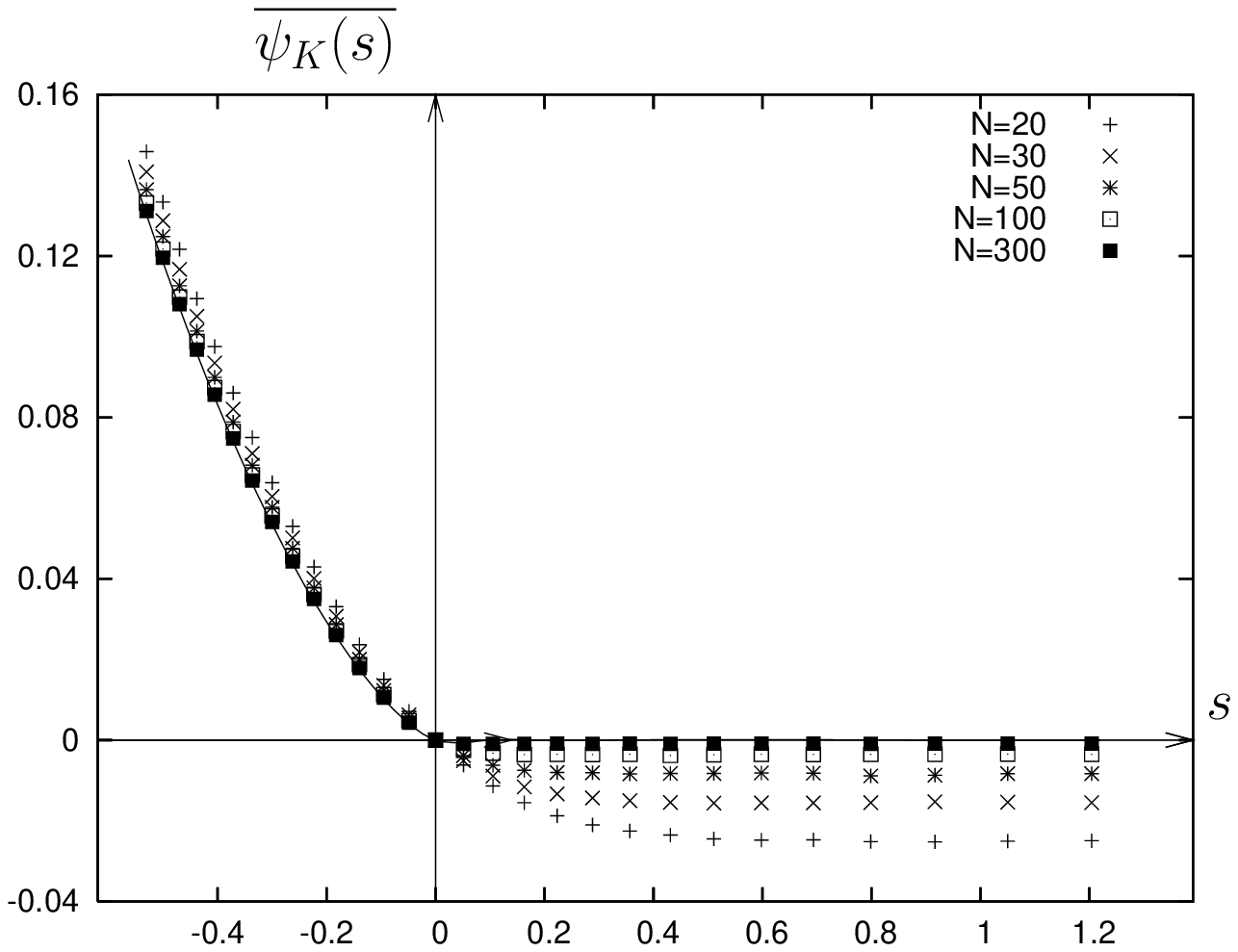}
\caption{Plot of $\overline{\psi_K(s)}$ as a
function of $s$ for different values of $N$ and $\mu=0.8$. The solid line is
the solution of Eq. (\ref{exactmean_trap}): for $s>0$ the curve approaches the
horizontal axis as $N^{-1/\mu}$ for $N\gg 1$.}
\label{fig1}
\end{centering}
\end{figure}

We have checked the presence of this dynamic transition by solving
numerically the eigenvalue equation (\ref{eig}). In Fig. \ref{fig1}, we show a
plot of $\overline{\psi_K(s)}$ as a 
function of $s$ for different values of $N = 20, 30, 50$ and
$100$ for $\mu=0.8$. The data were obtained by averaging over $10^6$
samples. For $s>0$ these
numerical data show that $\overline{ 
  \psi_K(s) } \to 0$ when $N \to \infty$ (one finds indeed $\overline{ 
  \psi_K(s) } \propto -N^{-1/\mu}$, consistent with the bounds in
Eq. (\ref{bounds_trap})). On the other hand, this plot on
Fig. \ref{fig1} shows that
$\lim_{N\rightarrow\infty}\overline{\psi_K(s) } > 0$ for $s<0$. The
solid line in Fig. \ref{fig1} 
is the analytical value of $\overline{\psi_K(s)}$ obtained by solving
numerically Eq.~(\ref{exactmean_trap}), which is in good agreement with the
numerical data and shows the presence of a dynamical phase transition. 

The slope of the large deviation $\overline{\psi_K(s)}$ function is related by
Eqs. (\ref{kmoy},\ref{def_psi}) to the mean value of the activity $\langle K
\rangle$ over all histories, thus in the active phase ($s<0$) of the dynamics
the activity takes a constant value $\langle K
\rangle>0$ while in the inactive phase ($s>0$) it is found to be
subextensive in the system size ($\langle K\rangle\sim N^{-1/\mu}$). This
coexistence of active and inactive phases of space-time, or {\it dynamical
  heterogeneity} is argued to manifest itself through the glassy properties of 
the directed Trap Model. In the corresponding pure system, where the waiting
times 
are no longer random variables, glassiness obviously disappears, and so does
the discontinuity in the derivatives of the large deviation function. The
activity 
is a constant $\langle K
\rangle>0$ also in the $s>0$ phase and no dynamical heterogeneity is present.

$\;$

\begin{figure}[h]
\begin{center}
\includegraphics[width=7.9cm]{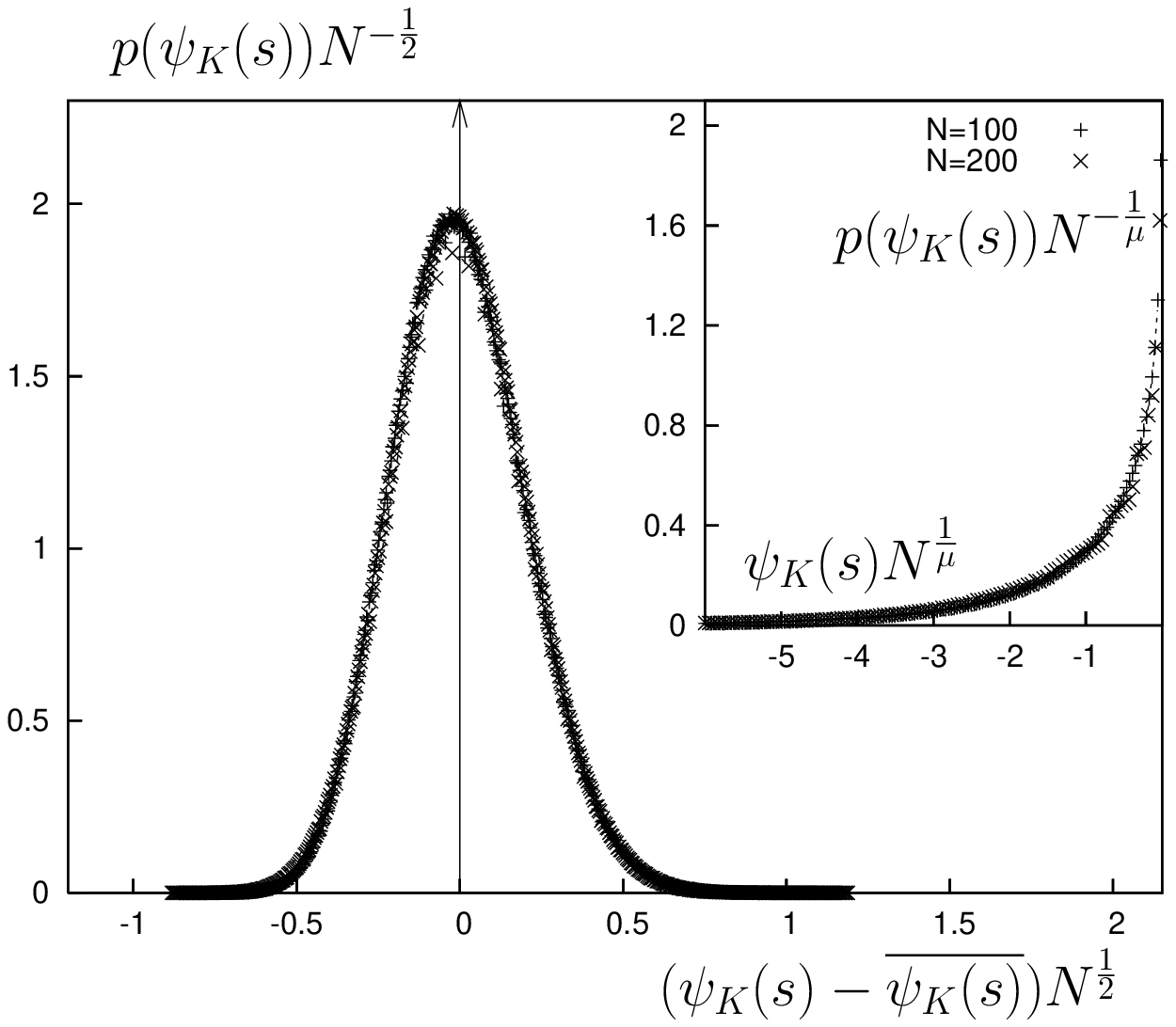}
\caption{Numerical evaluation of the distribution of the large
  deviation function for $\mu=0.8$ and 
  $s=-0.4$ (N=100,200). The solid line is the rescaling of the Gaussian distribution (\ref{dis_lambda}). {Inset}: numerical evaluation of the
  distribution of the large deviation function for $\mu=0.8$ and $s=0.2$. The
  solid line is the large $N$ limit of the rescaling of distribution (\ref{distr_spos}).}
\label{fig2}
\end{center}
\end{figure}

The numerical results in Fig. \ref{fig1} were obtained assuming that the large
deviation function approaches its average in the large $N$ limit and is distributed
according to Eqs. (\ref{dis_lambda},\ref{var}).
Indeed this supposition was confirmed evaluating numerically the probability
distribution $p(\psi_K(s))$ of $\psi_K(s)$ for a negative $s$. In
Fig. \ref{fig2} we show a plot of $p(\psi_K(s)) N^{-1/2}$ as a function
of $N^{1/2} (\psi_K(s) -  \overline{\psi_K(s)} )$ for $\mu = 0.8$ and $s=-0.4$
($N=100, 200$). The data were obtained by averaging over $3\cdot 10^7$
samples. The solid line is the Gaussian distribution
obtained rescaling by $\sqrt{N}$ the Eqs. (\ref{dis_lambda}, \ref{var}). This
gaussian id approached by the numerical data as the system size grows.  

On the other hand, for $s>0$, the evaluation of the average large deviation function, i.e. the
solution of Eq. (\ref{exactmean_trap}), was done limiting the
values of $\psi_K(s)$ to the $N$-dependent bound (\ref{bounds_trap}). Indeed
we found numerically that the distribution $p(\psi_K(s))$ of $\psi_K(s)$ for a
positive $s$ is well fitted by the form (\ref{max_dis}):
\be\label{distr_spos}
p(\psi_K(s))=N \mu\psi_K(s)^{\mu-1}(1-\psi_K(s)^{\mu})^{N-1} \;
\ee
for all $N$.
In the inset of Fig. \ref{fig2} we show the rescaled distribution
$p(\psi_K(s))N^{-1/\mu}$ as a function of $\psi_K(s)N^{1/\mu}$ for
$N=100,200$ averaged over $10^7$ samples for $s=0.2$ and $\mu=0.8$. These
distribution are found to be well approached by the large $N$ limit of
the rescaled distribution which takes the form $p(x)=\mu x^{\mu-1}\exp{(-x^\mu)}$. 

\section{Random Energy Model (REM)}

We consider now a widely explored model of disordered system, the Random
Energy Model introduced in \cite{derrida} as a mean field spin-glass
model. Its dynamics has been studied in detail by Koper and
Hilhorst in Ref. \cite{hilh}.

\subsection{Dynamics of the REM}

The Random Energy Model is a system of $2^\Nm$ energy
levels $E_i$ which are independent random variables distributed according to a
Gaussian distribution of zero mean and variance $\frac{1}{2}\Nm J^2$:
\be
P(E)=\frac{e^{-E^2/\Nm J^2}}{\sqrt{\pi \Nm J^2}}\; .
\ee
The model exhibits a phase transition at the critical temperature
$T_c=J/2\sqrt{\ln 2}$ below which the free energy becomes a constant
$E_0(\Nm)\simeq \Nm J \sqrt{\ln 2}$ in the thermodynamic limit and the
entropy vanishes. As a consequence, the energy levels close to $E_0(\Nm)$ will
dominate the low temperature 
phase. We thus consider a system of $N$ energy levels 
\be\label{low}
E_i=E_0(\Nm)+\ep_i \;, 
\ee
where $\ep_i$ is small and nonextensive \cite{hilh} distributed according
to
$$
p(\ep)=\left\{ 
\begin{array}{ll}
\rho\, e^{\rho (\ep-\ep_c)} , & \ep \geq \ep_c \;, \\
0\,,\quad & \ep<\ep_c \;,
\end{array}
\right.
$$
where $\rho=T_c^{-1}$ is a constant and $\ep_c$ is a cut-off energy. In the
following, we will consider the scaling limit $N\rightarrow \infty$,
$\ep_c\rightarrow \infty$, while $N e^{-\rho \ep_c}=v$ is kept fixed, where 
the physical properties of the REM for temperatures $T<T_c$ take finite,
cut-off independent values. 

The dynamics of this model is defined by a Master Equation 
\be\label{Masterrem1}
\frac{d P_i(t)}{dt}=\sum_{j\neq i}W_{ij}P_j(t)-\sum_{j\neq i}W_{ji}P_i(t) \;,
\ee
where the transition rates $W_{ji}$ for going from level $i$ to level 
$j$ are taken of the form \cite{hilh}:
\be\label{tr}
W_{ji}=B_j V_j V_i \qquad \text{with} \quad B_j=e^{-\beta
  \ep_j}\,, 
\ee
where $ i=1,\dots ,N$ and $V_i$ are positive random variables. One can easily
check that the rates $W_{ji}$ obey the detailed balance condition. 
The distribution of the $B_i$'s can be easily derived from the distribution of
the energies $\ep_i$, leading to:   
\begin{eqnarray}\label{dist_b_rem}
p(B)= 
\begin{cases}
\frac{v}{N} B^{-1-\mu} , &  \text{for} \quad
\left(\frac{v}{N}\right)^{1/\mu}<B \;,\\ 
0\,,\quad &  \text{otherwise}
\end{cases}
\end{eqnarray}
where we set $\mu=\rho/\beta=T/T_c$, $\mu\in (0,1)$. Following
Ref. \cite{hilh}, we will consider level dependent barriers $V_i = B_i^{-q}$
with $q\in(0,1)$. If one associates to each level $i$ an independent random
magnetization $\mu_i$ with zero mean and variance $N$, the kinetic REM can be
considered as a magnetic model. Within this choice of level dependent
barriers it has been shown that the behavior of
the equilibrium autocorrelation function for the magnetization exhibits a
non-exponential behavior \cite{hilh}
\begin{eqnarray}\label{auto}
\overline{C(t)}&\propto&  t^{-2\mu/q} \hspace*{0.8cm}\text{for}\; q+\mu>1 \,, \\
\overline{C(t)}&\propto& t^{-\eta}e^{-Dt^\gamma}
\hspace*{0.35cm}\text{for}\;
q+\mu<1 \;\text{and} \; q<\frac{1}{2} \,, \nonumber\\
\overline{C(t)}&\propto&  t^{-2\mu/(2q-1)}\; \text{for}\;  q+\mu<1
\;  \text{and} \; q>\frac{1}{2} \,,\nonumber 
\end{eqnarray}
where $D$ is a constant, $\gamma=\mu/(1-2q+\mu q/(1-q))$ and
$\eta=(2\mu/(1-q)-1)\gamma$. 

Here we apply  the thermodynamic formalism of histories to this model
and show that a dynamic transition takes place reflecting
the slow dynamics in this system (\ref{auto}).

\subsection{Dynamic transition}

To investigate the presence of a dynamic transition, we look
again at the large deviation function $\psi_K(s)$ defined above in
Eq. (\ref{def_psi}), associated to the number of configuration changes
$K(t)$ between time $0$ and time $t$. From the Master Equation governing the
dynamics (\ref{Masterrem1}), one obtains the equation of evolution of
$\tilde{P}_i(s,t)$ defined in Eq.~(\ref{def_ptilde}) 
\begin{align}
&\frac{d\tilde{P}_i(s,t)}{dt}=\sum_j\hat{W}^K_{ij}\tilde{P}_j(s,t)\\
&=e^{-s} B_iV_i\sum_{j\neq i}V_j \tilde{P}_j(s,t)-V_i\tilde{P}_i(s,t)\sum_{j\neq i}B_jV_j\,.\nonumber
\end{align}
The large deviation function $\psi_K(s)$ is the largest eigenvalue of the
evolution operator $\hat{W}^K$. The eigenvalues $\lambda_j(s)$ of $\hat{W}^K$
are solutions of the equation
\begin{align}
&f(\lambda_j(s)) = 1 \;,\label{eigen2} \\
&f(\la)=\sum_i \frac{e^{-s} B_i V_i^2}{\zeta V_i
  +(e^{-s}-1)B_iV_i^2 +\la}
\end{align}
where $\psi_K(s) = {\rm max}_{1 \leq j \leq N} \,\{\lambda_j(s)\}$ and $\zeta=\sum_iB_iV_i$.

We first focus on the case $q+\mu > 1$. One notices that $f(\la)$ has $N$
simple poles on the negative real axis at 
$x_i=-(\zeta V_i +(e^{-s}-1)B_iV_i^2 )<0$, and decreases to zero for 
$\la \rightarrow \infty$. One has also $f(0) = 1$ for $s=0$, whereas $f(0) >
1$ for $s>0$ and $f(0)<1$ if $s<0$. Thus in the active phase, $s<0$, there is
only one positive eigenvalue $\lambda_j$ which coincides with $\psi_K(s) >
0$. As done previously for the directed trap model, we thus suppose that
$\psi_K(s)$ is self averaging in the limit $N \to \infty$ and coincides with
its average. One obtains that, for $s<0$ 
\begin{eqnarray}\label{psi_sneg_rem}
\overline{\psi_K(s)} \sim N^{(2q+\mu-1)/\mu} \tilde
\psi_K(s) \;,
\end{eqnarray}
where $\tilde \psi_K(s)$ is independent of $N$ and solution of the equation
\be\label{psi_sneg_rem2}
\mu v \int_{v^{1/\mu}}^\infty \frac{B^{-\mu-2q}}{\tilde v B^{-q} +
  \tilde \psi_K(s)} 
dB = 
e^{s} \;,
\ee 
where $\tilde v = \mu v^{(1-q)/\mu}/(q+\mu-1) > 0$. Performing an analysis
similar to the one done for the directed trap model, one also finds that the
fluctuations of $\psi_K(s)$ around its mean value $\overline{\psi_K(s)}$ are
described by a Gaussian as in Eq. (\ref{dis_lambda}) of variance $N^{-1/2}$. 

However, the analysis in the inactive phase is more subtle. Indeed, given the
singularities of $f(\lambda)$ in $x_i$, together with the fact that $f(0) >
1$ for $s>0$, $\psi_K(s)$ satisfies the bounds 
\be\label{ineq1}
-\min_i(\zeta V_i +(e^{-s}-1)B_iV_i^2 )<\psi_K(s)< 0\;.
\ee
If one labels the random variables $B_i$ such that $B_1 < B_2 < ... < B_{N}$
one easily sees that the minimum in the left hand side of the inequality
(\ref{ineq1}) is reached for the maximum $B_N$, whose distribution is given in
the large $N$ limit by
\be\label{dist_max_rem}
p(B_N)=v \mu B_N^{-1-\mu} e^{-v B_N^{-\mu}} \;.
\ee
Using this distribution (\ref{dist_max_rem}) together with the distribution of
the $B_i$'s in Eq. (\ref{dist_b_rem}) one obtains that 
\be\label{low_bound}
-c_1 N^{(q+\mu-1)/\mu} <\overline{\psi_K(s)}< 0 \;,
\ee
where $c_1 > 0$ is a constant, independent of $N$. 
To find an upper bound for $\overline{\psi_K(s)}$, one writes the equation
above (\ref{eigen2}) in a different form:
\be\label{eigen3}
\sum_i \frac{e^{-s} B_i V_i}{e^{-s}\zeta  +(1-e^{-s})(\zeta-B_iV_i +\la(s)V_i^{-1})}=1\;. \nonumber
\ee
Since the first term in the denominator is the sum of the numerators, the
remaining terms $(1-e^{-s})(\zeta-B_iV_i +\la(s)V_i^{-1})$ must be negative for
at least one $i$. In this way we get an upper bound for $\psi_K(s)$:
\begin{equation}\label{bounds1}
\psi_K(s)<(e^{-s}-1)\min_i(\zeta V_i-B_iV_i^2)
\end{equation}
Again, the minimum in the right hand side of this inequality is reached for
the maximum $B_N$. Performing the average, one gets
\be\label{up_bound}
\overline{\psi_K(s)}< -c_2 N^{(q+\mu-1)/\mu} \;.
\ee
where $0< c_2 < c_1 $ is a constant. Combining these two bounds
(\ref{low_bound}, \ref{up_bound}) one gets for $s>0$
\be\label{psi_spos_rem}
-\overline{\psi_K(s)} =  {\cal O}(N^{(q+\mu-1)/\mu}) \;.
\ee 
Thus by looking at the behavior of $\overline{\psi_K(s)}$ for large
$N$ in Eq. (\ref{psi_sneg_rem}, \ref{psi_sneg_rem2}, \ref{psi_spos_rem}), one 
shows that there is a dynamical transition as $s$ crosses
$0$. Eq. (\ref{psi_sneg_rem2}) shows that $\tilde \psi_K(s) \propto
(-s)^{q/(q+\mu-1)}$ for small $s$ indicating that, as in the directed
trap model, the order of this dynamical transition is larger than one.

\begin{figure}[h!]
\begin{centering}
\includegraphics[width=7.7cm]{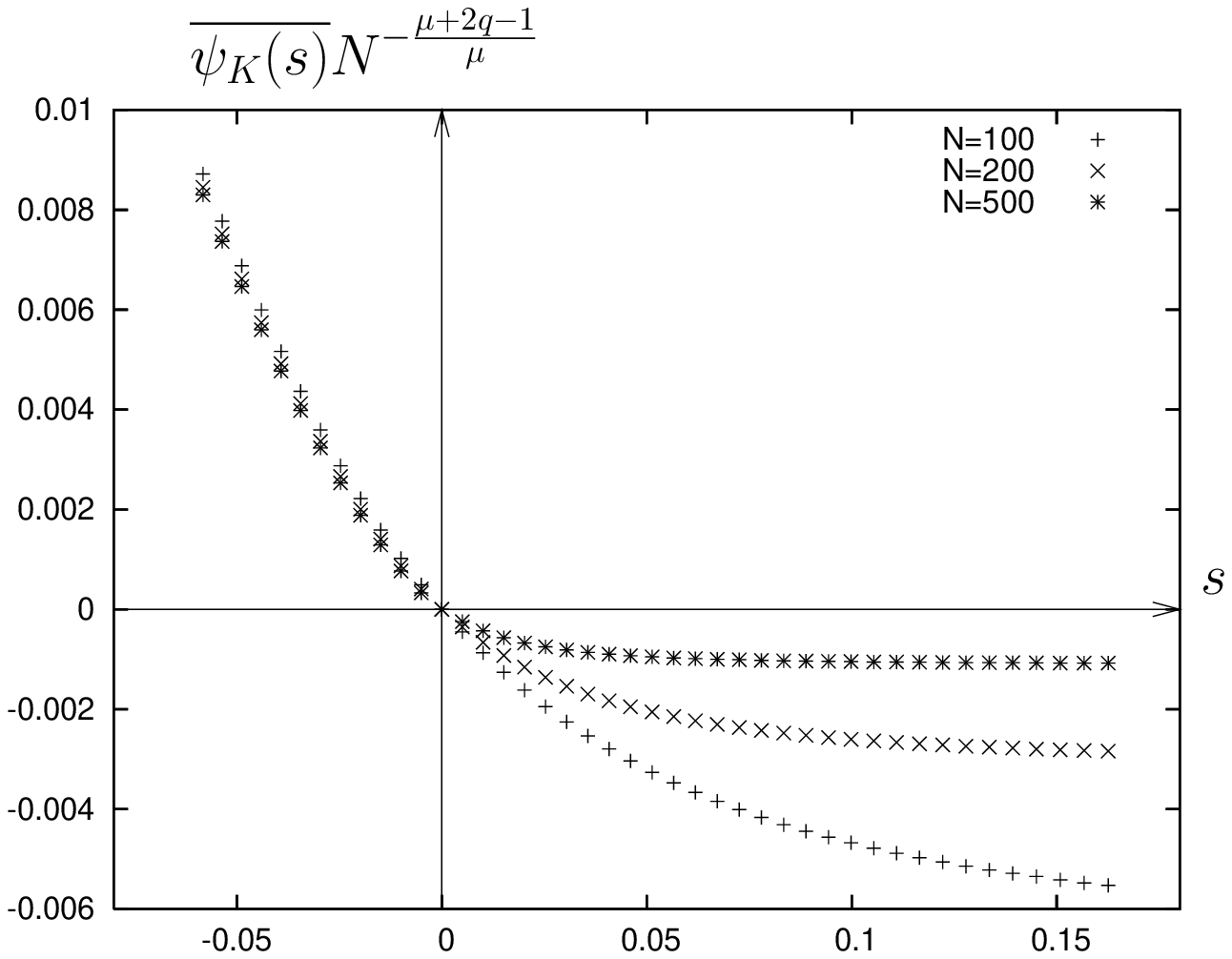}
\caption{Numerical evaluation of the large deviation function $\psi_K(s)$ for $q=0.9$ and $\mu=0.8$ ($q+\mu>1$) in the Random Energy Model.}
\label{qmumag1}
\end{centering}
\end{figure}

These analytical predictions for $q+\mu >1$ in Eq. (\ref{psi_sneg_rem},
\ref{psi_spos_rem}) have been confirmed by solving
numerically the eigenvalue equation (\ref{eigen2}). The average value
$\overline{\psi_K(s)}$ was computed by 
averaging over at least $10^6$ realizations of the disorder. In
Fig. \ref{qmumag1}, 
we show a plot of $N^{-(\mu+2q-1)/\mu} \overline{\psi_K(s)}$ as a
function of $s$. In agreement with
Eq. (\ref{psi_sneg_rem},~\ref{psi_spos_rem}), one observes a dynamical
transition 
occurring at $s=0$. Moreover it has been checked that for positive $s$ the
large deviation function behaves like predicted in Eq. (\ref{psi_spos_rem}),
{\it i.e.} $-\overline{\psi_K(s)} =  {\cal O}(N^{(q+\mu-1)/\mu})$.

\begin{figure}[h!]
\begin{centering}
\includegraphics[width=7.7cm]{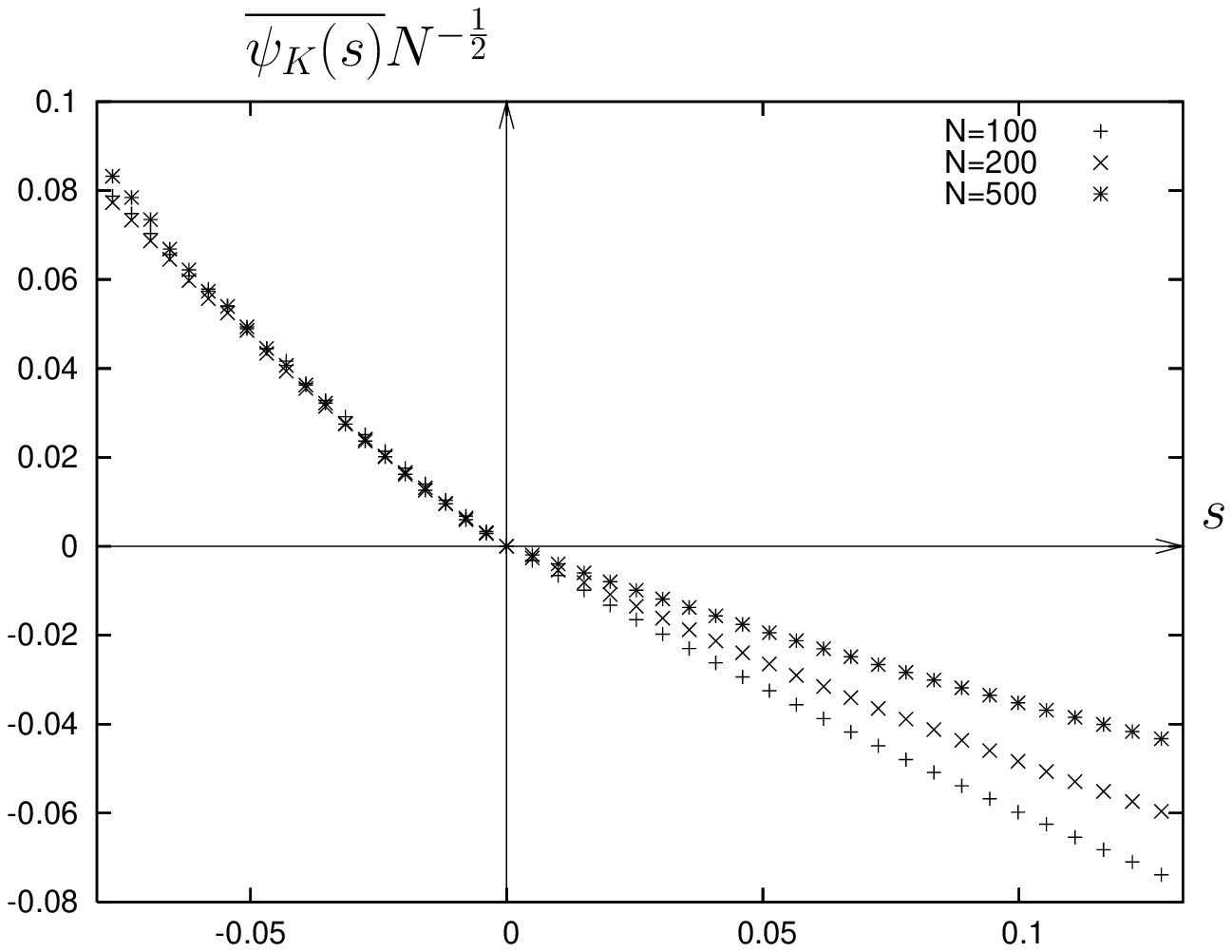}
\caption{Numerical evaluation of the large deviation function for $q=0.45$ and
  $\mu=0.45$ ($q+\mu<1$ and $q<1/2$) in the Random Energy Model.}
\label{qmumin1a}
\end{centering}
\end{figure}

\begin{figure}[h!]
\begin{centering}
\includegraphics[width=7.7cm]{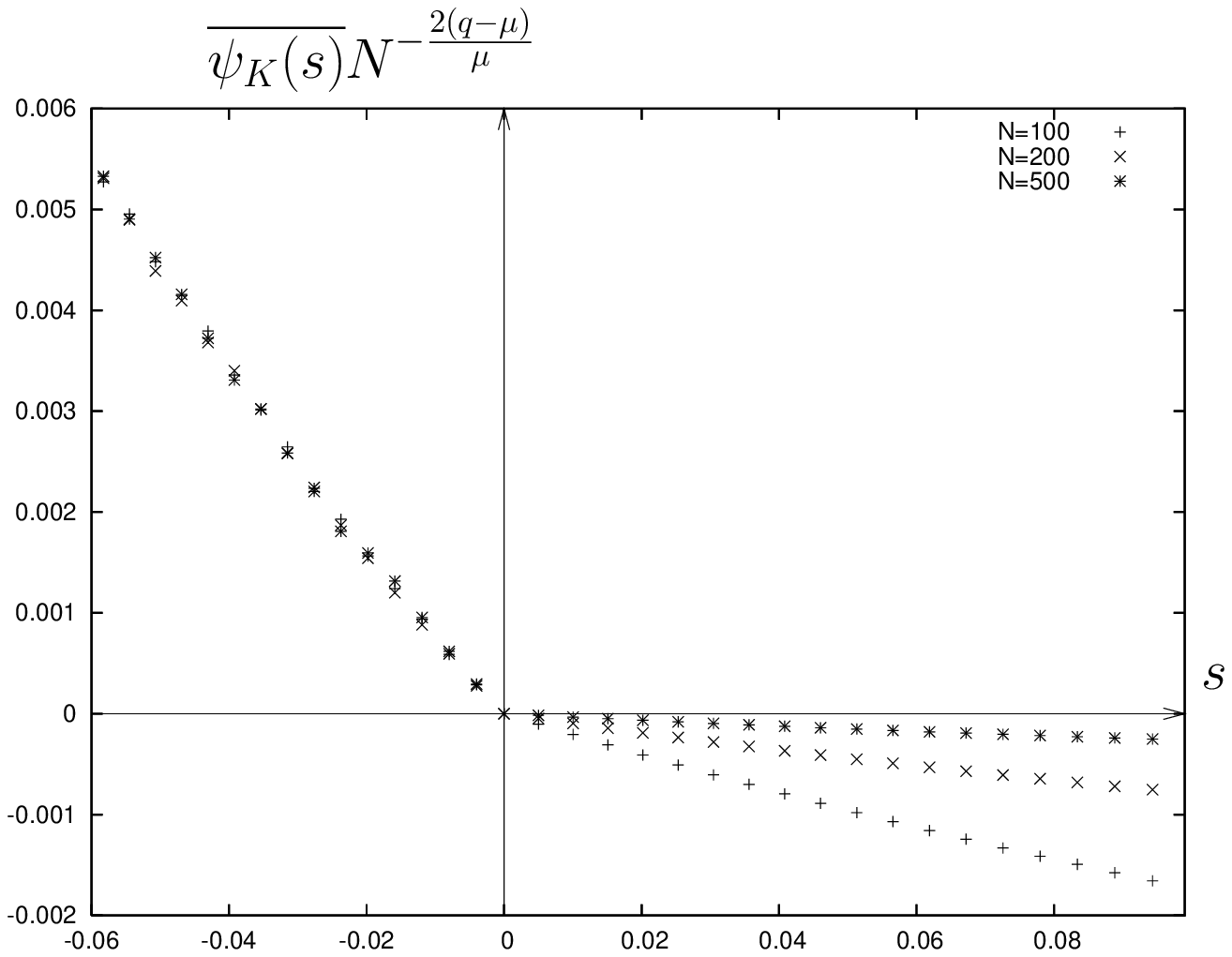}
\caption{Numerical evaluation of the large deviation function for $q=0.6$ and
  $\mu=0.35$ ($q+\mu<1$ and $q>1/2$) in the Random Energy Model.}
\label{qmumin1b}
\end{centering}
\end{figure}

We have checked numerically that this transition is also present for
$q+\mu <1$, where the equilibrium correlation function displays a
non-exponential decay (\ref{auto}). In Fig. \ref{qmumin1a}, one shows
a plot of $N^{-1/2}\overline{\psi_K(s)}$ as a function of $s$ for
$q=0.45$ and $\mu=0.45$. In that regime, the autocorrelation function
decays like a stretched exponential and our numerical data clearly
shows a transition occurring for $s=0$. Similarly, in
Fig. \ref{qmumin1b}, one shows
a plot of $\overline{\psi_K(s)}N^{-2(q-\mu)/\mu}$ as a function of $s$ for
$q=0.6$ and $\mu=0.35$, where the correlation function decays like a
power law. Here again, our numerical data clearly
shows a dynamical transition occurring for $s=0$. Again these results were
obtained averaging over at least $10^6$ realizations of the disorder. 
Thus in the three cases we have an anomalous behavior of the
correlation function (\ref{auto}) and a discontinuity in the
derivatives of the large 
deviation function $\overline{\psi_K(s)}$ at $s=0$
(Fig. \ref{qmumag1}-\ref{qmumin1b}).

\section{Connection with the extreme value statistics}

The glassy behavior of disordered systems emerges in the low
temperature phase or in the long time limit. Consequently the extreme values of
the disorder realizations play a dominating role with respect to the typical
values. The low temperature phase of the Random Energy Model is indeed
described through the distribution of the lowest energy levels (\ref{low}),
which is found to belong to the Gumbel universality class of the minimum of
variables which are unbounded and have a distribution that decays faster than
any power law to $-\infty$.

It was pointed out in \cite{extreme,extr} that the Gumbel class corresponds
exactly 
to the one step replica symmetry breaking solution in the replica language. It
was also highlighted that for the Weibull class, {\it i.e.} the extreme value
statistics 
of bounded variables, and the Fr\'echet class for power-law decaying variables,
the replica method cannot be used without substantial modifications.

In order to shed light on how the extreme value statistics of disorder
affects the glassy properties of the system in the low temperature phase we
consider two versions of the Random Energy Model. We choose for
the equilibrium distribution of configurations in the low temperature phase, i.e. the Boltzmann factors (\ref{tr}) $B_j~=~e^{-\beta\ep_j}$, both an exponential and a uniform
distribution. We find a dynamical transition in the first case
but not in the second. These results are corroborated by a study of the
equilibrium autocorrelation function in both cases. 

\subsection{Exponentially distributed Boltzmann factors}

We recall that the dynamics of the Random
Energy Model can be described through the following Master equation: 
\be\label{Masterrem2}
\frac{d P_i(t)}{dt}=\sum_{j\neq i}W_{ij}P_j(t)-\sum_{j\neq i}W_{ji}P_i(t)\;,
\ee
where the transition rates for level-dependent barriers are
$W_{ji}=N^{-2p}B_j^{1-q}B_i^{-q}$, with $p\geq 0$ and $q\in(0,1)$.

We define as a version of the Random Energy Model a system whose dynamics is
still described by Eq. (\ref{Masterrem2}), with the difference that the
statistics of the energy barriers in the low temperature phase has
changed. Instead of a L\'evy distribution we consider a probability distribution
function of the Boltzmann factors of the form:
\be
p(B_i)=p(e^{-\beta \ep_i})=N\theta(B_i) e^{-NB_i}\;,
\ee
where the scaling in $N$ has been chosen {\it a posteriori} to find a well
defined thermodynamic limit of the equilibrium autocorrelation function (see
section \ref{correl_fun_N}). 

\begin{figure}[h]
\begin{centering}
\includegraphics[width=7.7cm]{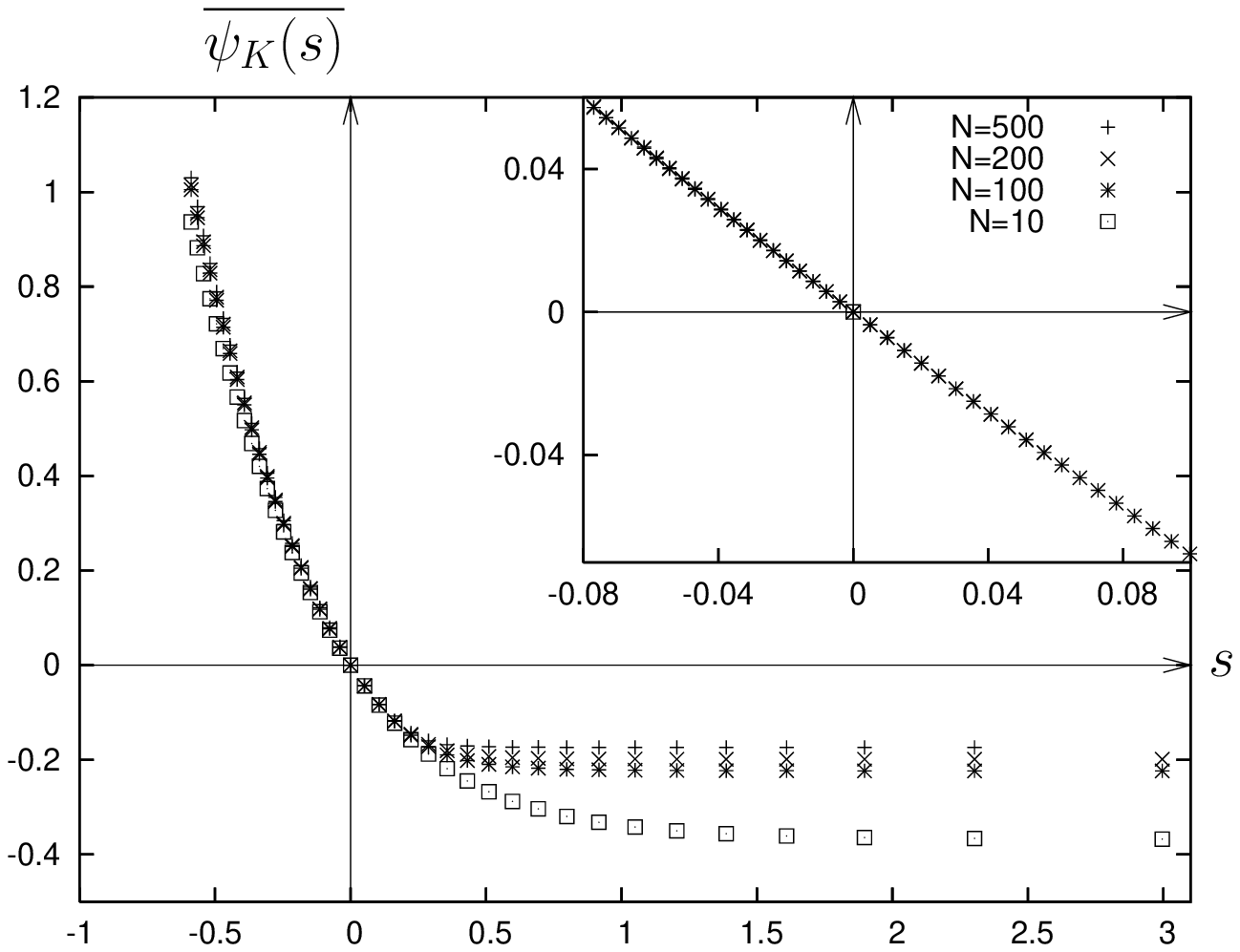}
\caption{Numerical evaluation of the average large deviation function $\overline{\psi_K(s)}$ for exponentially distributed Boltzmann factors, with $p=q=0.9$. Inset: numerical evaluation of $\overline{\psi_K(s)}$ for uniformly distributed Boltzmann factors.}
\label{exp}
\end{centering}
\end{figure}

The large deviation function $\psi_K(s)$ is again the largest solution of the eigenvalue
equation
\be\label{eigen9}
\sum_i \frac{e^{-s} B_i^{1-2q}}{\zeta B_i^{-q} +(e^{-s}-1)B_i^{1-2q} +\psi_K(s)}=1\;.
\ee
We solved numerically the equation (\ref{eigen9}) and averaged the solution
over $10^7$ realizations of the disorder. The results are plotted in
fig. \ref{exp}.  

As in the directed Trap Model and the Random Energy Model we observe a
dynamical transition. For $s<0$ the large deviation function appears to be
constant in the system size. For $s>0$ the curve approaches the
horizontal axis for $N\gg 1$.

Thus even if the moments of the Boltzmann factors (or the mean trapping times)
are well defined, the system presents dynamical heterogeneities and we
expect an anomalous slowing down of the dynamics.

\subsection{Uniformly distributed waiting times} 

Let's consider uniformly distributed Boltzmann factors:
\be
p(B_i)=N\theta\left(B_i-\frac{a}{N}\right)\theta\left(\frac{a+1}{N}-B_i\right)
\;, 
\ee 
where $a$ is a finite constant with $0\notin [a,a+1]$. If we allow the $B_i$'s
take the value $B=0$, after some time the system will be frozen in a single
energy level.

Evaluatig numerically the large deviation function of this system we find a complete independence of the large deviation function from the
system size for all $s$ (see the inset in fig. \ref{exp}), and no discontinuity
of the large deviation function. Since we do not observe any dynamic transition
in the space of histories, we expect this system to be dynamically
homogeneous.

\subsection{Correlation function}

To compute the equilibrium autocorrelation
function of the Random Energy Model with
exponentially and uniformly
distributed Boltzmann factors we recast the Koper and Hilhorst's study \cite{hilh} of the relaxation properties of the Random Energy Model.  

Starting from the Master Equation (\ref{Masterrem2}) with transition rates
(\ref{tr}), the authors proved that the autocorrelation function $C(t)$ of any physical
quantity that does not couple to the energies of the REM can be written as a
complex integral of the form:
\be\label{5.3}
C(t)=-\oint \frac{ds}{2\pi i}\frac{e^{-st\zeta}\sum_{j=1}^N \sum_{k\neq j}^N \frac{V_jB_j V_k B_k}{(s-V_j \zeta)(s-V_k\zeta)}}{s Z\sum_{l=1}^N \frac{ V_l B_l}{s-V_l\zeta}}
\ee
with $Z = \sum_{i=1}^N B_i$ and where the line integral is taken over a
contour oriented counterclockwise and 
encloses all the poles of the integrand on the positive real axis. 
If we average this expression, as in \cite{hilh}, over a L\'evy distribution
we find the asymptotic behaviors (\ref{auto}) which characterize the
glassiness of the Random Energy Model.

Here, to corroborate the presence or absence of the dynamical transition
obtained previously, we
average Eq. (\ref{5.3}) over an exponential distribution and
then over a uniform distribution.

\subsubsection{Exponential distribution}\label{correl_fun_N}

We consider exponentially distributed Boltzmann factors
\be
p(B)=N\theta(B) e^{-NB}\;,
\ee
and transition rates of the form
\be
W_{ij}=N^{-2q}B_i^{1-q}B_j^{-q}\;,
\ee
where the $N$ dependence is chosen to give a well defined thermodynamic
limit of the correlation function. After some manipulations which details are
given in Appendix \ref{appendix}, one finds
\be\label{correl_exp_text}
\overline{C}(t)=t^\frac{4}{q+1}\int_0^\infty dx\, \int_0^\infty dy\,
xy\,e^{-t^{\frac{1}{q+1}}\tilde g(x,y)} \;,
\ee 
where $\tilde g(x,y)=x+y+\Gamma(2-q)/(x^q+y^q)$. One easily finds that $\tilde
g(x,y)$ has a single minimum in $\mathbb{R}^+ \times \mathbb{R}^+$ for 
${x}^*={y}^*=(q\Gamma(2-q)/4)^{1/(q+1)}$ and
$\partial^2_x g({x}^*,{y}^*)=\partial^2_y g(x^*,y^*)={x^*}^{-1}$. Thus the
large time behavior $\overline{C}(t)$ is given by a saddle point calculation
yielding 
\be
\overline{C}(t)\sim 2\pi\, t^\frac{3}{q+1}\,{x^*}^3
e^{-t^\frac{1}{q+1}\tilde g(x^*,y^*)}\quad\text{for}\; t \gg 1 
\ee
which shows, given that $q > 0$, that the relaxation is indeed slower than
exponential.

\subsubsection{Uniform distribution}

We now consider uniformly distributed Boltzmann factors:
\be
p(B)=N\theta(B-\frac{a}{N})\theta(\frac{a+1}{N}-B)\;,
\ee
where $a$ is a constant. After some manipulations left in the
Appendix \ref{appendix}, one has
\be\label{correl_uni_text}
\overline{C}(t)=\int_a^{a+1}dx\int_a^{a+1}dy
\frac{xy}{a^2}e^{-t\frac{a^{1-q}}{x^q+y^q}} \;.
\ee
The minimum of the integrand is reached for $x=y=a$ and the maximum for
$x=y=a+1$, thus the integral is bounded from above and below by an exponential
and we conclude that:
\be\label{uni_corr}
\overline{C}(t)\sim e^{-tc(a)}\;,
\ee
where $c(a)$ is a positive function of $a$.
Eq. (\ref{uni_corr}) is the asymptotic behavior of the correlation
function of a non-glassy system. 

\section{Conclusion}

The central statement of our work is that when a system presents
glassy features (anomalous slowing down of the dynamics, non exponential 
decay of the correlation functions, anomalous dependence of viscosity
on temperature and so on) it appears heterogeneous in time and energy landscape,
and a dynamical transition in the space of time realizations the
system can follow in configuration space has to be found. 
 
Here we validated this statement for two well known disordered models of
glasses, the Random Energy Model and the Directed Trap Model, detecting in
both a dynamical phase transition. 
In the Random Energy Model the influence of the statistics of disorder over
the space-time heterogeneity of the system was analyzed. Indeed we
found a dynamical transition for a REM with exponentially distributed
Boltzmann factors and showed that we are in presence of a glassy dynamics
by computing the equilibrium autocorrelation function of this system
which is shown to decay anomalously, {\it i.e.} to be a stretched
exponential in time. Furthermore we considered a uniform distribution
of Boltzmann factors finding no dynamical transition and showing, as
expected, that the equilibrium relaxation is exponential in time and
no slowing down of the dynamics is present. 

\section*{ACKNOWLEDGMENTS}

We thank C. Monthus for several useful discussions. This work was supported by
the French Ministry of Education through Grant No. ANR-05-JCJC-44482.
 
\begin{appendix}

\section{computation of the correlation functions for the REM}\label{appendix}   

In this appendix, we present some details concerning the computation of the
corrrelation function for the kinetic REM. 

\subsubsection{General framework}

We recall the Master Equation (\ref{Masterrem2}) with transition rates
(\ref{tr}): 
\be\label{mgen}
\frac{dP_i(t)}{dt}= V_i B_i\sum_{j=1}^N V_j{P_j(t)}-V_i P_{i}(t)\zeta\;.
\ee
The probability to find the system at level $i$ at time $t$ if it was in $j$
at time $t=0$, {\it i.e.} the Green function, can be written in terms of the
eigenvalues and eigenfunctions of the Master Equation.
We call $\phi_i^\la$ the $i$th component of the right-hand eigenfunction with
eigenvalue $-\la$:
\be
\frac{d \phi_i^\la}{dt}=-\la \phi_i^\la\;.
\ee
Using (\ref{mgen}) one can write:
\be\label{teng}
\phi_i^\la=\frac{V_i B_i}{V_i \zeta-\la}\sum_{j=1}^NV_j\phi_j^\la\;.
\ee
The detailed balance condition implies that the left-hand eigenfunction is
$\psi_i^\la=\phi_i^\la/B_i$, so that the Green function takes the form:
\begin{align}\label{gh}
G_{ij}(t)&=\sum_{l=1}^N e^{-\la_l t}\frac{\phi_i^{\la_l} \psi_j^{\la_l}
}{\sum_{k=1}^N\phi_k^{\la_l} \psi_k^{\la_l}}\nonumber \\
&=\sum_{l=1}^N e^{-\la_l t}\frac{\frac{V_i B_i}{V_i \zeta-\la_l}\frac{V_j}{V_j \zeta-\la_l}}{\sum_{k=1}^N\frac{V_k^2 B_k}{(V_k \zeta-\la_l)^2}}\;.
\end{align}

The equilibrium autocorrelation function expressed in terms of these Green
functions is 
\be\label{green}
C(t)\equiv \sum_{j=1}^N \{G_{jj}(t)- G_{jj}(\infty)\} P_j^{\text{eq}}\;,
\ee
where $P_j^{\text{eq}} = \lim_{t \to \infty}P_j(t) \propto
B_j$. 

To get rid of the dependence of the autocorrelation function $C(t)$ in
Eq.~(\ref{green}) on the unknown 
eigenvalues $\la_l$ we consider its Laplace transform $\tilde{C}(s)$, 
which using (\ref{gh}) can be written~as:
\be
\tilde{C}(s)=\frac{1}{Z}\sum_{l=2}^{N}\frac{H(\la_l)}{(s+\la_l)F'(\la_l)}
\ee
with $Z = \sum_{i=1}^N B_i$ and where
\be
H(z)=\sum_{j=1}^N\frac{V_j^2 B_j^2}{(V_j \zeta-z)^2} \; , \;
F(z)=\sum_{j=1}^N\frac{V_j^2 B_j}{V_j\zeta-z}-1\;. \nonumber
\ee
We introduce $g(z) = H(z)/((s+z)F(z))$, 
which has poles in $z=-s$, in $z=V_j\zeta$ and in the zeroes of
$F(z)$. Looking at the eigenvalue equation (\ref{teng}) we see that the sum of
the 
residues of $g(z)$ in the zeroes of $F(z)$ will be exactly $\tilde{C}(s)$.
Thus, since the contour integral of $g(z)$ along a circle centered in the
origin with radius $R$ vanishes as $R\rightarrow \infty$, one finds:
\begin{widetext}
\be
\tilde{C}(s)=-\frac{1}{s}\left\{ \frac{\sum_{j= 1}^N \sum_{k \neq j}^N \frac{V_j
        B_j V_k B_k}{(V_j \zeta+s)(V_k\zeta+s)}}{Z\sum_{l=1}^N \frac{V_l
        B_l}{V_l\zeta+s}} +\frac{1}{Z^2}\sum_{j=1}^N B_j^2
    -1\right\}\;,
\ee
\end{widetext}
from which we see that $\tilde{C}(s)$ has poles only on the negative real
axis ($s=0$ is not a pole). Applying an Inverse Laplace transform and changing
$s$ into $-s \zeta$ we obtain the expression of
the autocorrelation function $C(t)$ given in the text in Eq. (\ref{5.3}). 

\subsubsection{Exponential distribution}

We consider exponentially distributed Boltzmann factors
\be
p(B)=N\theta(B) e^{-NB}\;,
\ee
Using for the denominator in Eq. (\ref{5.3}) the integral
representation $
\alpha^{-1}=\int_0^\infty d\la e^{-\alpha\la}$, 
we obtain from Eq. (\ref{5.3}) the expression for the averaged autocorrelation
function:  
\begin{widetext}
\begin{eqnarray}
&&\overline{C}(t)=\lim_{N\rightarrow \infty} N^4\int_0^\infty dB_i\int_0^\infty
dB_j\int_0^\infty d\la \int_0^\infty d\mu \oint \frac{ds}{2\pi i}\frac{B_i
  B_j}{s(sB^q_i-1)(sB^q_j-1)}    e^{-N(B_i+B_j)} \label{formula_C}\\
&& \times \, 
e^{-\la
  (B_i+B_j)-\mu(\frac{B_i}{sB^q_i-1}+\frac{B_j}{sB^q_j-1})-stN^{-2q}(B_i^{1-q}+B_j^{1-q})    } \, \times \, I_N^{N-2} \nonumber
\end{eqnarray}
\end{widetext}
where the integral $I_N$ is given by
\be
I_N=N \int_0^\infty dB e^{-NB-\la B-\mu
  \frac{B}{sB^q-1}-st{N^{-2q}}B^{1-q}}\,. \nonumber
\ee
Rescaling the variable $B=B'/N$ and integrating by parts we have, to leading
order in $N$: 
\be\label{I_largeN}
I_N\simeq 1-\frac{1}{N}\int_0^\infty dB e^{-B}(\la-\mu+stN^{-q}(1-q)B^{-q})
\ee
Thus using this large $N$ expansion (\ref{I_largeN}) in the formula
(\ref{formula_C}), and integrating over $\la$ and $\mu$ we get:
%
\begin{widetext}
\begin{equation}
\overline{C}(t)=\lim_{N\rightarrow \infty} N^4 \int_0^\infty dB_i
\int_0^\infty dB_j
\oint\frac{ds}{2\pi i}\frac{B_iB_j}{s(B_i+B_j+1)}
\frac{e^{-N(B_i+B_j)-stN^{-2q}(B_i^{1-q}+B_j^{1-q}+N^{q}\Gamma(2-q))}}{s(B_i^qB_j+B_iB_j^q)-B_i-B_j-(sB^q_i-1)(sB^q_j-1)} \nonumber
\end{equation}
\end{widetext}
Changing variables according to $NB_i=x$, $NB_j=y$ and keeping only the
leading orders in $N$ we have:
\begin{align}
&\overline{C}(t)=\lim_{N\rightarrow \infty} N^4 \int_0^\infty dx
\int_{0}^\infty dy
\oint\frac{ds}{2\pi i} x y \nonumber \\
& \times \,\frac{e^{-(x+y)-stN^{-q}\Gamma(2-q)}}{s(sN^{-q}(x^q+y^q)-1)}\;,
\end{align}
which has one pole in $s=N^q/(x^q+y^q)$.
Complex integration thus gives:
\be
\overline{C}(t)=\int_{0}^\infty dx \int_{0}^\infty dy \,
xye^{-(x+y)}e^{-t\frac{\Gamma(2-q)}{x^q+y^q}}\ee 
Finally we change variables $x=x't^{1/(q+1)}$,
$y=y't^{1/(q+1)}$ and find the expression given in the
text~(\ref{correl_exp_text}).

\subsubsection{Uniform distribution}

We now consider uniformly distributed Boltzmann factors:
\be
p(B)=N\theta(B-\frac{a}{N})\theta(\frac{a+1}{N}-B)\;,
\ee
where $a$ is a constant.
The average of the correlation function (\ref{5.3}) takes the form:
\begin{widetext}
\begin{eqnarray}
&&\overline{C}(t)=\lim_{N\rightarrow
  \infty}N^4 \int_{\tfrac{a}{N}}^{\tfrac{a+1}{N}}dB_i
\int_{\tfrac{a}{N}}^{\tfrac{a+1}{N}}dB_j \int_0^\infty d\la \int_0^\infty d\mu
\oint 
\frac{ds}{2\pi i} \frac{B_iB_j}{s(sB^q_i-1)(sB^q_j-1)} \nonumber \\
&& \times \, e^{-\la
  (B_i+B_j)-\mu(\frac{B_i}{sB^q_i-1}+\frac{B_j}{sB^q_j-1})-stN^{-2p}(B_i^{1-q}+B_j^{1-q})} J_N^{N-2} 
\end{eqnarray}
\end{widetext}
where
\be\label{ii}
J_N= \int_{\tfrac{a}{N}}^{\tfrac{a+1}{N}} dB N e^{-\la B-\mu\frac{B}{sB^q-1}-stN^{-2p}B^{1-q}}\;.
\ee
Since the interval of integration is very small when $N\gg 1$ we can consider
the integrand as a constant over this interval and make the approximation:
\be
J_N \simeq e^{-\la{a}{N^{-1}}+\mu{a}{N^{-1}}-stN^{-2p-(1-q)}a^{1-q}}
\ee
Thus after integration over $\la$ and $\mu$ and changes of variables $x=NB_i$
and $y=NB_j$, the correlation function becomes:
\begin{eqnarray}
&&\overline{C}(t)= \lim_{N\rightarrow \infty}  
\int_a^{a+1}dx \int_a^{a+1}
dy \oint\frac{ds}{2\pi i} xy \nonumber \\
&&\times \frac{e^{-stN^{-2p+q}a^{1-q}}}{sa^2(sN^{-q}(x^q+y^q)-1)} 
\end{eqnarray}
The integrand has one pole in $s=N^q/(x^q +y^q)$ so after complex integration
and choosing $p=q$ we obtain the formula given in the text
(\ref{correl_uni_text}).

\end{appendix}

\end{document}